\documentclass[runningheads]{llncs}
\usepackage[utf8x]{inputenc}
\usepackage[T1]{fontenc}

\usepackage{amsmath}
\usepackage{amsfonts}
\usepackage{amssymb}
\usepackage{stmaryrd}
\usepackage{latexsym}
\usepackage{cmll}
\usepackage{graphicx}
\usepackage{hyperref}

\usepackage[dvipsnames]{xcolor}
\definecolor{aawhite}{rgb}{0.97,0.97,0.97}
\definecolor{awhite}{rgb}{0.90,0.90,0.90}
\definecolor{lgreen}{rgb}{0.94,1.0,0.98}
\definecolor{dgreen}{rgb}{0.0,0.3,0.1}
\definecolor{sgreen}{rgb}{0.0,0.7,0.3}
\definecolor{lgreen}{rgb}{0.94,1.0,0.98}
\definecolor{bgreen}{rgb}{0.00,0.50,0.25}
\definecolor{dblue}{rgb}{0.0,0.1,0.6}
\definecolor{lblue}{rgb}{0.8,0.8,1.0}
\definecolor{mixed}{rgb}{0.0,0.3,0.3}
\definecolor{dred}{rgb}{0.6,0.2,0.0}
\definecolor{sred}{rgb}{0.7,0.2,0.0}
\definecolor{ddred}{rgb}{0.3,0.1,0.0}
\definecolor{turq}{rgb}{0.28,0.82,0.80}
\definecolor{lyellow}{rgb}{1.00,0.97,0.94}
\definecolor{mygreen}{rgb}{0,0.6,0}
\definecolor{mygray}{rgb}{0.5,0.5,0.5}
\definecolor{mymauve}{rgb}{0.58,0,0.82}
\definecolor{codegreen}{rgb}{0,0.6,0}
\definecolor{codegray}{rgb}{0.5,0.5,0.5}
\definecolor{codepurple}{rgb}{0.58,0,0.82}
\definecolor{backcolour}{rgb}{0.95,0.95,0.90}
\definecolor{eclipseOrange}{RGB}{200,48,0}
\definecolor{eclipseBlue}{RGB}{0,0,172}

\RequirePackage[textsize=footnotesize,shadow,color=white,linecolor=orange]{todonotes}
\RequirePackage[normalem]{ulem}

\usepackage{listings}
\renewcommand{\verb}{\lstinline}
\lstdefinelanguage{yarel} {
mathescape=true,
texcl=false,
keywords={iter, where, times, var, int, bool, module, import, dcl, def, dec, id, inc, neg, !dec, !id, !inc, !neg, if, then, else, elif, return, fix, !, it, rec},
morekeywords={package, public, class, static, void, new, throws, for, interface, private, final },
literate=	
	{0}{{\textcolor{orange}{0}}}1
	{1}{{\textcolor{orange}{1}}}1
	{2}{{\textcolor{orange}{2}}}1
	{3}{{\textcolor{orange}{3}}}1
	{4}{{\textcolor{orange}{4}}}1
	{5}{{\textcolor{orange}{5}}}1
	{6}{{\textcolor{orange}{6}}}1
	{7}{{\textcolor{orange}{7}}}1
	{8}{{\textcolor{orange}{8}}}1
	{9}{{\textcolor{orange}{9}}}1
	,
    morecomment=[l]{//}, 
    morecomment=[s]{/*}{*/}, 
    morestring=[s]{\\}{\\} 
}

\lstdefinestyle{yarel-style}{
	backgroundcolor=\color{backcolour},
	commentstyle=\color{codegreen},
	keywordstyle=\color{magenta},
	numberstyle=\tiny\color{codegray},
	stringstyle=\color{codepurple},
	basicstyle=\ttfamily\footnotesize,
	breakatwhitespace=false,
	breaklines=true,
	captionpos=b,
	keepspaces=true,
	numbers=left,
	numbersep=5pt,
	showspaces=false,
	showstringspaces=false,
	showtabs=false,
	tabsize=2
}
\lstdefinelanguage{java}{
    morekeywords = [1]{import, abstract, class, enum, extends
                      , implements, import, instanceof, interface, native
                      , new, final,  package, private, protected, public
                      , static, void},
    morekeywords = [2]{boolean, int, do, for, if, else, throws, catch, while
                      , try, null, length, assert, case, return, super, this},
    morekeywords = [3]{ },
    morekeywords = [4]{ },
    morekeywords = [5]{ },
    keywordstyle = [1]\color{magenta},
    keywordstyle = [2]\color{blue},
    keywordstyle = [3]\color{magenta},
    keywordstyle = [4]\color{orange},
    keywordstyle = [5]\color{lblue},
    sensitive = true,
    morecomment = [l]{//},
    morecomment = [s]{/*}{*/},
    morecomment = [s]{/**}{*/},
    commentstyle={\color{dgreen}},
    morestring = [b]",
    morestring = [b]',
    basicstyle={\small\ttfamily\bfseries}, 
    stringstyle={\ttfamily\small\color{orange}},
    numbers=left,
    numberstyle=\tiny\color{mygray},
    xrightmargin=0em,
    xleftmargin=3em,
    stepnumber=1,
    numbersep=1em,
    lineskip=-0.5ex,
    mathescape=true,
    showstringspaces=false,
    frame=none,
    breaklines=true,
    columns=[l]{fullflexible},
    keepspaces=true,
}

\lstnewenvironment{java}{\lstset{language={java},}}{}

\lstdefinestyle{java-style}{
	backgroundcolor=\color{backcolour},
	commentstyle=\color{codegreen},
	keywordstyle=\color{magenta},
	numberstyle=\tiny\color{codegray},
	stringstyle=\color{codepurple},
	basicstyle=\fontsize{9}{13}\selectfont\ttfamily,
	breakatwhitespace=false,
	breaklines=true,
	captionpos=b,
	keepspaces=true,
	numbers=left,
	numbersep=5pt,
	showspaces=false,
	showstringspaces=false,
	showtabs=false,
	tabsize=2
}
\newcommand{\vj}[1]{\lstinline[language=java,style=java-style]+#1+}

\usepackage{caption}
\DeclareCaptionFont{white}{\color{white}}
\DeclareCaptionFormat{listing}{\colorbox[cmyk]{0.43, 0.35, 0.35,0.01}{\parbox{\textwidth}{\hspace{15pt}#1#2#3}}}
\begin{document}
\title{Interleaving classical and reversible}
%
%
\author{Armando B. Matos\inst{1} \and
	Luca Paolini\inst{2}\orcidID{0000-0002-4126-0170} \and
	Luca Roversi\inst{2}\orcidID{0000-0002-1871-6109}}
\authorrunning{A. Matos, L. Paolini, L. Roversi}
%
\institute{Universidade do Porto, Departamento de Ciência de Computadores\\
	\email{armandobcm@yahoo.com}\\
	\and
	Università degli Studi di Torino, Dipartimento di Informatica, Italy\\
	\email{\{luca.paolini,luca.roversi\}@unito.it}}

\maketitle
\begin{abstract}
Given a simple recursive function, we show how to extract two interacting processes from it.
The two processes can be described by means of iterative programs, one of which is intrinsically reversible, in a language that, up to minor details, belongs to the core of widely used imperative programming languages.
We implement the two processes as interleaving synchronous
\textsf{JAVA} threads whose interaction is equivalent to the recursive function they are extracted from.
\end{abstract}

\section{Introduction}
\label{section:Introduction}

Typically, scientific works on reversible computations promote reversibility as an interesting topic, for many situations exist in which a computational activity has to be able to retrace its steps; \cite[Part I]{perumalla2013chc} is a standard reference to a first list of those situations.
As a reinforcement, we think that knowing how to program in a reversible way,
over time, could become a natural mental scheme, analogous to
recursive and iterative programming schemes, when the computational problem to be solved contemplates its necessity. This work contributes to such an idea.

We deal with reversible computing in the lines of those
techniques that \cite[Chapter 9]{perumalla2013chc} identifies as ``Adding Reversibility to Irreversible Programs''.

Mainly, in this work, we add reversibility to irreversible programs
according to a basic scheme that we can intuitively describe
in some steps.
Given a classical recursive definition, we extract from it iterative reversible and classical parts that we can make to collaborate according to a synchronous Producer/Consumer pattern in order to implement the initially given recursion. Using Perumalla's terminology \cite[Chapter 9]{perumalla2013chc}, it turns out that we can identify reversible aspects in a classical computation by means of a method that sits in between
``Source-to-Source Translator'' and ``Library-Based'' approaches.

In order to drive the intuition, let \vj{recF[p,b,h]}
(see \textbf{Listing~\ref{primitive-recursive-function}})
be a recursive function defined in a reasonable
	programming formalism
	on top of a \emph{predecessor} function \vj{p}, a \emph{step} function \vj{h}, and a \emph{base} function \vj{b}. We decompose \vj{recF[p,b,h]}
          into two processes \vj{itFCls[b,h]} and \vj{itFRev[p,pInv]} such that:
\begin{align}
\label{align:compilation scheme}
\mbox{\vj{recF[p,b,h]}} & \simeq
\mbox{\vj{itFCls[b,h]}} \parallel  \mbox{\vj{itFRev[p,pInv]}}
\enspace,
\end{align}
where
(i) ``$ \simeq $'' stands for ``\emph{equivalent to}'',
and ``$ \parallel $'' for ``\emph{interaction/parallel}-composi\-tion''
between its two arguments;
(ii) \mbox{\vj{itFCls[b,h]}} sequentially composes \vj{b} and a
\vj{for}-loop that iteratively applies \vj{h};
(iii) \vj{itFRev[p,pInc]} sequentially composes two \vj{for}-loops one iterating \vj{b}, the other its inverse \vj{bInv}.

On one side,  \eqref{align:compilation scheme} says that we can
translate the classical language construct \vj{recF[p,b,h]} as
composition of two interacting parts, one of which naturally exposes a reversible
nature; so we see \eqref{align:compilation scheme} as
instance of ``Source-to-Source Translator'' approach.
On the other, concerning the ``Library-Based'' approach, \eqref{align:compilation scheme}  allows to look at
the reversible part \vj{itFRev[p,pInv]} as the component of a library, with only reversible code in it, that produces the values that
the classical part relies on to accomplish the overall task
to implement \vj{recF[p,b,h]} together with \vj{itFCls[b,h]}.

In order to realize the Producer/Consumer pattern that
\eqref{align:compilation scheme} summarizes, we use a
programming syntax that, up to minor syntactic details,
is a nucleus common to \textsf{C}-style programming languages (nowadays ubiquitous, e.g.  \textsf{C\raisebox{.1ex}{++}}, \textsf{JAVA}, \textsf{C\#}, \ldots ) which can be implemented by
the control structures of the two reversible languages \textsf{SRL}
\cite{DBLP:journals/tcs/Matos03,MatosRC2020} and \textsf{RPP} \cite{paolini2017ngc,DBLP:journals/tcs/PaoliniPR20,MatosRC2020}.
We recall that \textsf{SRL} and \textsf{RPP} are intrinsically
reversible programming notations with finite iterations only,
as expressive as the class of primitive recursive functions.
This essential correspondence
allows us to implement the Producer/Consumer
pattern in terms of \textsf{JAVA} synchronous threads
that the interested reader can experiment with \cite{JAVAcode}.

The reminder of this introduction deepens, but at an intuitive level, the observations
that, starting from the idealized formalization in \eqref{align:compilation scheme} eventually leads to the \textsf{JAVA} prototypical implementation in \cite{JAVAcode}.

\begin{figure}
\begin{lstlisting}[
language = java,
style = java-style,
caption = The recursive function \vj{recF}. ,
label = primitive-recursive-function
]
 Fix recF(x)                         {
     if   (x==0) { b(x);           }
     else        { h(x,recF(x-1)); } }
\end{lstlisting}
\end{figure}

To start with, let us focus on the recursive \vj{recF} of \textbf{Listing~\ref{primitive-recursive-function}} which is one
of the possible instances of \vj{recF[p,b,h]} in \eqref{align:compilation scheme},
where \vj{b(x)} is the \emph{base} function, \vj{h(x,y)} the \emph{step} function,
and \vj{p(x)} the \emph{predecessor} \vj{x-1};
moreover, the \emph{condition} \vj{c(x)}, which identifies the base case, is set to \vj{x==0}.
As an (almost) minor remark, some of the readers would identify \vj{recF}
as primitive recursive; others would say it is not exactly as such because its
inductive step has not form \vj{h(p(x),f(p(x)))}. Of course, we can always
to think of taking an \vj{h(x,y)} that hides \vj{p(x)} applied
to its first argument.

\begin{figure}
\begin{lstlisting}[
language = java,
style = java-style,
caption = Iterative unfolding \vj{recF(3)}: the bottom-up part. ,
label = iteratve simulation of f(3) recursive
]
  /*** Assumption: the inital value of x is 3 */
  x = p(x)       // ==2
  x = p(x)       // ==1
  x = p(x)       // ==0
  y = b(x)       // ==b(p(p(p(3))))
  y = h(x,y)     // ==h(p(p(p(3))),b(p(p(p(3)))))
  x = pInv(x)    // ==pInv(p(p(p(3))))==p(p(3))
  y = h(x,y)     // ==h(p(p(3)),h(p(p(p(3))),b(p(p(p(3))))))
  x = pInv(x)    // ==pInv(p(p(3)))==p(3)
  y = h(x,y)     // ==h(p(3),h(p(p(3))
                 //         ,h(p(p(p(3))),b(p(p(p(3)))))))
  x = pInv(x)    // ==pInv(p(3))==3
  y = h(x,y)     // ==h(3,h(p(3),h(p(p(3))
                 //      ,h(p(p(p(3))),b(p(p(p(3))))))))
\end{lstlisting}
\end{figure}

\textbf{Listing~\ref{iteratve simulation of f(3) recursive}}
reconstructs the unfolding of \vj{recF(3)},
i.e.~\mbox{\vj{h(3,h(p(3),h(p(p(3)),}} \mbox{\vj{h(p(p(p(3))),b(p(p(p(3))))))))}}.
Every of its comments asserts a property of the values that \vj{x} or \vj{y} stores.
Lines 2--4 unfold an iteration that computes \vj{p(p(p(3)))}, which eventually sets the value of \vj{x} to \vj{0}.
Line 5 starts the construction of the final value of \vj{recF(3)} by
applying the base case of \vj{recF}, i.e.~\vj{b(x)}.
By definition, let \vj{pInv} denote the inverse of \vj{p},
i.e.~\vj{pInv(p(z))==p(pInv(z))==z}, for any \vj{z}.
Clearly, in our running example, the function \vj{pInv(x)} is \vj{x\+1}.
Lines 6--13 alternate \vj{h(x,y)}, whose result \vj{y}, step by step, gets better and better the final value \vj{recF(3)}, and \vj{pInv(x)},
which produces a new value for \vj{x}.

\begin{figure}
\begin{lstlisting}[
	language = java,
	style = java-style,
	caption = Iterative \vj{itF} equivalent to \vj{recF}.,
	label = iteratve version of recF
	]
	/*** Assumptions. s == 0, e == 0, g == 0, w == 0 */
	w = w + x;
	for (i = 0; i<=w; i++)      {
		if      (x> 0) { g++; }
		else if (x==0) { e++; }
		else           { s++; }
		x = p(x);               }

	for (i = 0; i<=w; i++)                {
		x = pInv(x);
		if      (x> 0) { g--; y = h(x,y); }
		else if (x==0) { e--; y = b(x);   }
		else           { s--;             } }
	w = w - x;
\end{lstlisting}
\end{figure}

Let us call \vj{itF} the code in \textbf{Listing}~\ref{iteratve version of recF}.
It implements \vj{recF} by means of finite iterations only.
Continuing with our running example, if we run \vj{itF} here above starting with
\vj{x == 3}, then \vj{x == 0} holds at line 8, just after the first \vj{for}-loop;
after the second \vj{for}-loop \vj{y == recF(3)} holds at line 14.

\begin{figure}
\begin{lstlisting}[
language = java,
   style = java-style,
caption = Reversible side of \vj{itF}.,
label = reversible core itF
]
 /*** Assumptions. s = 0, e = 0, g = 0, w = 0 */
 w = w + x;
 for (i=0; i<=w; i++)      {
   if      (x> 0) { g++; } //number of times x is `g'reater than 0
   else if (x==0) { e++; } //number of times x is `e'qual to 0
   else           { s++; } //number of times x is `s'maller than 0
   x = p(x);               }

 for (i=0; i<=w; i++)                                           {
   x = pInv(x);
   if      (x> 0) { g--; /* Value of x for h availabe here */ }
   else if (x==0) { e--; /* Value of x for b availabe here */ }
   else           { s--;                                      } }
 w = w - x;
\end{lstlisting}
\end{figure}

Let us call \emph{reversible side} of \vj{itF} the code in \textbf{Listing}~\ref{reversible core itF}, i.e.~it is \textbf{Listing}~\ref{iteratve version of recF} without
\vj{h(x,y)} and \vj{b(x)} at lines 11, 12. It is reversible
because, when it stops,
the value that every variable contains, but \vj{i}, is identical to the one it contained at
line 2, independently from the assumptions at line 1.
The variable \vj{i} does not contradict that \textbf{Listing}~\ref{reversible core itF}
is reversible, for a finite iteration can hide it
\cite{paolini2017ngc,DBLP:journals/tcs/PaoliniPR20}.


The structure of the \emph{reversible side} \vj{itF} has two parts.
Through lines 2--7, the variable \vj{g} counts how many times \vj{x} remains positive, the variable \vj{e} how many it stays equal to \vj{0}, and the variable \vj{s} how many it becomes negative.
In our running example, it will never be the case that
\vj{x} assumes a negative value because the whole iteration at lines
3--7 is driven by the initial value of \vj{x} which, initially,  is assumed
to be non negative, and \vj{p(x)}
decreases its argument of a single unity.
We will discuss about the relevance of variable \vj{s} later.
Lines 9--13 simply undo what lines 2--7 do; the reason is that
the execution of \vj{p(x)}, \vj{g\+\+}, \vj{e\+\+}, \vj{s\+\+}
is annihilated by executing their inverses \vj{pInv(x)}, \vj{g--}, \vj{e--}, \vj{s--},
respectively, in reversed order.
This is why the correct values of \vj{x} are available
at lines 12, 11 and we can use them as arguments of
\vj{b(x)} and \vj{h(x,y)} in order to update \vj{y}
as in \textbf{Listing}~\ref{iteratve version of recF}, according to the results
we obtain by the recursive calls to \mbox{\vj{recF(x)}} in \textbf{Listing}~\ref{primitive-recursive-function}.

The reason not to call \vj{b(x)} and \vj{h(x,y)} at lines 12, 11 of \textbf{Listing}~\ref{reversible core itF}, as compared to
\textbf{Listing}~\ref{iteratve version of recF}, is that we want
\textbf{Listing}~\ref{reversible core itF} to be the \emph{reversible side}
of \vj{itF}; calling \vj{b(x)} and \vj{h(x,y)} in it would generate the
result \vj{y}, so preventing the possibility to reset the value of every
involved variable to their initial value.
This is why we also need a \emph{classical side} of \vj{itF} that
generates \vj{y} in collaboration with the \emph{reversible side} in order to
implement \vj{recF(x)} correctly.

We claim that \textbf{Listing}~\ref{loop on the classical side} and~\ref{loop on the reversible side} are
the \emph{classical side} \vj{itFCls} and
the \emph{reversible side} \vj{itFCRev} of \vj{itF}
which, suitably interacting, make the scheme \eqref{align:compilation scheme}
almost fully concrete.
We see \vj{itFRev} as the producer of values that the consumer \vj{itFCls} consumes as soon as it uses them as arguments of the \emph{base} \vj{b(x)} and the \emph{step} function \vj{h(x,y)}.
The following points illustrate how \vj{itFCls} and \vj{itFRev} synchronously interact.

\begin{enumerate}
	\item The starting point of the synchronous interaction
	between \vj{itFCls} and \vj{itFRev} is line 2 of \vj{itFCls}.
	The comment:
	\begin{align*}
	&\mbox{\vj{/* Inject the current x at line 2 of itFRev to let it start */}}
    \end{align*}
	describes what, in a fully implemented version of
	\vj{itFCls}, we expect in that line of code.
	The comment says that \vj{itFCls} injects (sends, puts)
	its input value \vj{x} to line 2 of the
	\emph{reversible side} \vj{itFRev} (cf. \textbf{Listing}~\ref{loop on the reversible side}).
    Once \vj{itFRev} obtains that
	value at its line 2, as outlined by the comment:
	\begin{align*}
	&\mbox{\vj{/* Inject here the value of x from line 2 of itFCls */}}
    \end{align*}
	its \vj{for}-loop at lines 4--8 is executed.

	\item After executing its line 2, \vj{itFCls} stops at line 3. It waits for \vj{itFRev}
to produce the number of times that \vj{itFCls} has to iterate
line 7. The comment:
\begin{align*}
	&\mbox{\vj{/*  Probe line 9 of itFRev to get the number of iterations to execute */}}
\end{align*}
says that we see the consumer as probing the producer \vj{itFRev}
until it delivers that value; \vj{itFRev} let that value be available in its variable \vj{g} at line 9, as outlined by comment:
\begin{align*}
	&\mbox{\vj{/* itFCls probes here g which has the number of iterations */}}
	\enspace .
\end{align*}

	\item Once gotten the value in \vj{iterations},
\vj{itFCls} proceeds to line 5 and
stops again. It waits for \vj{itFRev} to produce the
argument of \vj{b} which is eventually available for
probing at line 14 of \vj{itFRev}.

\item Once the argument becomes eventually available, then \vj{b} is applied,
and \vj{itFCls} enters its \vj{for}-loop, stopping at line 7 at every iteration.
The reason is that \vj{itFCls} waits for line 12 in \vj{itFRev} to produce the value of
the first argument of \vj{h(x,y)}. This interleaved dialog between line 7 of \vj{itFCls}
and line 12 of \vj{itFRev} lasts \vj{iterations} times.
\end{enumerate}

Identifying the \emph{reversible} side \vj{itFRev} and the \emph{classical} side \vj{itFCls}
of the classical recursive \vj{recF} in accordance with \eqref{align:compilation scheme}
recalls Girard's decomposition $ A \rightarrow B \simeq\ !A \multimap B$ \cite{GIRARD19871}
which looks at a classical computation
$ A \rightarrow B $ as an interaction between a linear (ideally
reversible) part with type $ C \multimap B $, for some $ C $, and a
non-linear $ !A $ that replicates computational resources;
conclusions will discuss this briefly again.

\begin{figure}
\begin{lstlisting}[
language = java,
style = java-style,
caption = Classical side: the consumer \vj{itFCls} to implement \vj{itF}. ,
label = loop on the classical side
]
  /*** Assumption. The value of the input x is available here */
  /* Inject the current x at line 2 of itFRev to let it start */
  iterations = /* Probe line 9 of itFRev to get the
                  number of iterations to execute   */
  y = b(/* Probe line 14 of itFRev to get the argument */);
  for (i = 0; i<iterations; i++)                  {
    y = h(/* Probe line 12 itFRev to get
             the first argument of h     */ , y); }
\end{lstlisting}
\end{figure}

\begin{figure}
\begin{lstlisting}[
language = java,
style = java-style,
caption = Reversible side \vj{itFRev}: the producer to implement \vj{itF}. ,
label = loop on the reversible side
]
  s = 0, e = 0, g = 0, w = 0;
  x = /* Inject here the value of x at line 2 of itFCls */
  w = w + x;
  for (i = 0; i<=w; i++)      {
    if      (x> 0) { g++; }
    else if (x==0) { e++; }
    else           { s++; }
    x = p(x);                 }
  /* itFCls probes here g which has the number of iterations */
  for (i = 0; i<=w; i++)                                    {
    x = pInv(x);
    if      (x> 0) { g--; /* itFCls probes here the
                             first argument value of h */ }
    else if (x==0) { e--; /* itFCls probes here the
                             argument value of b       */ }
    else           { s--;                                 } }
  w = w - x;
\end{lstlisting}
\end{figure}

\begin{figure}
\begin{lstlisting}[
language = java,
style = java-style,
caption = The generic structure of \vj{recG}.  ,
label = recG
]
  Fix recG(x)                      {
    if (x<=0) { b(x);            }
    else      { h(x,recG(p(x))); } }
\end{lstlisting}
\end{figure}

\paragraph{Contributions and road map.}
\begin{itemize}
\item
Let a \emph{predecessor} function \vj{p(x)} be given.
By definition, let $ \Delta_{\mbox{\vj{p}}} $ be defined as
\vj{p(x)-x} which, forcefully, is negative and which we assume constant.
Let also \vj{b(x)} and \vj{h(x,y)} be \emph{base} and \emph{step}
functions, respectively.
\textbf{Section}~\ref{section:From recursion to iteration}
identifies the structure of the \emph{reversible side}
that we can extract from any recursive classical function \vj{recG[p,b,h]},
built on \vj{b(x)}, \vj{h(x,y)}, \vj{p(x)},
and with a \emph{condition} \vj{c(x)} with form \vj{x <= 0}.
This means that \vj{recG} weakens the structure of \vj{recF} in \textbf{Listing}~\ref{primitive-recursive-function},
generalizing its use for programming.

\item
\textbf{Section}~\ref{section:Iteration as Producer/Consumer}
illustrates and implements the above Producer/Consumer pattern that corresponds to \eqref{align:compilation scheme} by means of \textsf{JAVA} classes.
The class that implements the producer has the structure of the
\emph{reversible side} that we will see in \textbf{Section}~\ref{section:From recursion to iteration};
the class that implements the producer synchronously interacts with the consumer,
whose structure \vj{itFCls} with minor structural updates.
Sources of the \textsf{JAVA} classes and packages are available at \cite{JAVAcode}.

\item \textbf{Section}~\ref{section:Conclusions} will address natural developments, open questions, and related works.
\end{itemize}


\section{From recursion to iteration}
\label{section:From recursion to iteration}
We show how, and justify why a recursive function \vj{recG}
slightly more general than \vj{recF} in \textbf{Listing}~\ref{primitive-recursive-function}
requires to generalize the structure of the \emph{reversible side}
we already commented in \textbf{Listing}~\ref{loop on the reversible side}.

The structure of \vj{recG} that we consider is in \textbf{Listing}~\ref{recG}
where \vj{b(x)} and \mbox{\vj{h(x,y)}} are a \emph{base} and a \emph{step}
function, respectively.
Let us recall from the introduction that, given a \emph{predecessor}
\vj{p(x)}, we let $ \Delta_{\mbox{\vj{p}}} $ be the
negative difference defined as $ \Delta_{\mbox{\vj{p}}} = \mbox{\vj{p(x)-x}}$.
The value of $ \Delta_{\mbox{\vj{p}}} $ that we consider is
an arbitrary and \emph{constant} $ k\ \mbox{\vj{ <= -1}} $, not only
$ k\ \mbox{\vj{ == -1}} $,
which requires to consider the slightly more general
\emph{condition} \vj{x <= 0}.
For example, let \vj{p(x)} be \vj{x-2}.
The computation of \vj{recG(3)} is
\vj{h(3,h(p(3),h(p(p(3)),b(p(p(3))))))}, which looks for the least
$ n $  of iterated applications of \vj{p(x)} such that
\vj{p(...p(3)...) <= 0} holds: in our case $\mbox{\vj{2 == }}\ n\ \mbox{\vj{< 3}} $.

We call \vj{itG} in \textbf{Listing}~\ref{iteratve recG}
the generalization of \vj{itF}
in \textbf{Listing}~\ref{iteratve version of recF}.
The scheme \vj{itG} iteratively implements any recursive function whose
structure can be brought back to the one of \vj{recG}.
Please, remark that line 1 initializes ancillae
\vj{s}, \vj{e}, \vj{g}, and \vj{w}, like for
the namesake variables of \vj{itF},
a standard initialization in reversible computing \cite{paolini2017ngc,perumalla2013chc}, and line 2 adds new ancillae \vj{z}, \vj{predDivX}, and \vj{predNotDivX}.

We also assume an initial \emph{non negative} value for \vj{x}.
The reason is twofold.
Firstly, it keeps our discussion as simple as possible,
with no need to use the absolute value of \vj{x} to set
the upper limit of every index \vj{i}
in the \vj{for}-loops tha occur in the code.
Second, negative values of \vj{x} would widen our discussion about
what a classical recursive function on negative values is and about
what its reversible equivalent iteration has to be; we see this as a
very interesting subject connected to \cite{BOITEN1992139}.

\begin{figure}
\begin{lstlisting}[
language = java,
style = java-style,
caption = The iterative reversible function \vj{itG} equivalent to \vj{recG}.,
label = iteratve recG
]
  s = 0, e = 0, g = 0, w = 0;
  z = 0, predDivX = 0, predNotDivX = 1;
  w = w + x; /* x is assumed to be the input */
  for (i = 0; i <= w; i++)   {
    if      (x >  0) { g++; }
    else if (x == 0) { e++; }
    else             { s++; }
    x = p(x);                  }

  for (i = 0; i < e; i++)                 {
    predDivX = predDivX + predNotDivX;
    predNotDivX = predDivX - predNotDivX; }

  for (j = 0; j < predDivX; j++)            {
  	for (i = 0; i <= w; i++)               {
  		x = pInv(x);
  		if      (x >  0) { g--; y = h(x,y); }
  		else if (x == 0) { e--; y = b(x);   }
  		else             { s--;             }}}

  for (j = 0; j < predNotDivX; j++)                           {
    w++;
    for (i = 0; i <= w; i++)                                 {
    	x = pInv(x);
    	if      (x >  0) { g--;
                         x = p(x);
                         if      (z <  0) {                }
                         else if (z == 0) { y = b(x); z++; }
                         else             { y = h(x,y);    }
                         x = pInv(x);                       }
    	else if (x == 0) { e--;             }
    	else             { s--;             }}
    w--;                                                      }
  w = w - x;
  /* y carries the output */
\end{lstlisting}
\end{figure}

We start observing that line 3 of \vj{itG} sets \vj{w} to the initial value
of \vj{x}; the reason is that every \vj{for}-loop, but the one at lines 10--13, has to last \vj{x\+1} iterations, and \vj{x} changes in the course of the computation; so, \vj{w} stores the initial value of \vj{x}
and stays constant from line 4 through line 21.
In fact it can change at lines 23--33. We will see why, but
eventually \vj{w} is reset to its initial value \vj{0} at line 34.

With the previous premises, given a non negative \vj{x}, in analogy to \vj{itF},
the \vj{for}-loop at lines 4--8 of \vj{itG}
iterates the application of \vj{p(x)} as many times as \vj{w\+1},
i.e.~the initial value of \vj{x} plus \vj{1}.
So, the value of \vj{x} at line 9 is equal to
$ \mbox{\vj{w\+(w\+1)*}}\Delta_{\mbox{\vj{p}}} $ which cannot be positive.
In particular, all the values that \vj{x} assumes in the \vj{for}-loop at
lines 4--8 belong to the following interval:
\begin{align}
	\label{align:interval of the iteration}
	I(\mbox{\vj{w}})
	& \triangleq [
	\mbox{\vj{w\+(w\+1)*}}\Delta_{\mbox{\vj{p}}}
	,\mbox{\vj{w\+w*}}\Delta_{\mbox{\vj{p}}}
	,\ldots
	,\mbox{\vj{w\+}}\Delta_{\mbox{\vj{p}}}
	,\mbox{\vj{w}}]
\end{align}
from the least to the greatest;
the counters \vj{g}, \vj{e}, \vj{s} say how many elements of
$ I(\mbox{\vj{x}}) $ are \vj{g}reater, \vj{e}qual or \vj{s}maller
than \vj{0}, respectively.
Depending on \vj{0} to belong to $ I(\mbox{\vj{x}}) $ determines the
behavior of the reminder part of \vj{itG}, i.e.~lines 10--34.

\par\vspace{\baselineskip}
We need to distinguish two cases in order to illustrate them.

\medskip
\emph{As a first case},
let $ \mbox{\vj{w\%}}\Delta_{\mbox{\vj{p}}}\ \mbox{\vj{== 0}} $,
i.e.~$ \Delta_{\mbox{\vj{p}}} $ divides the initial value of \vj{x}
which \vj{w} stores in it.
So, $ \mbox{\vj{0}} \in I(\mbox{\vj{x}}) $, which implies the following
relations hold at line 9:
\begin{align}
	\mbox{\vj{e == 1}}
	&&
	\mbox{\vj{g == -}} \frac{\mbox{\vj{w}}}
	                        {\Delta_{\mbox{\vj{p}}}}
	&&
	\mbox{\vj{s == (w\+1)-g-e}}
	\enspace.
\end{align}
Lines 10--12 execute exactly once, swapping \vj{predDivX} and \vj{predNotDivX}. We observe that we could have well used the
\vj{if}-selection in \textbf{Listing}~\ref{if alternative to for} here below,
which is a construct of \textsf{RPP}, in place of the \vj{for}-loop at lines 10--12; we opt for a more compact code with no empty branches.

\begin{lstlisting}[
language = java,
style = java-style,
caption = A possible replacement of lines 10--12 in \textbf{Listing}~\ref{iteratve recG}.,
label = if alternative to for
]
 if      (e <  0) {                                       }
 else if (e == 0) { predDivX = predDivX+predNotDivX;
                    predNotDivX = predDivX - predNotDivX; }
 else             {                                       }
\end{lstlisting}
\noindent
Swapping \vj{predDivX} and \vj{predNotDivX} let \vj{predDivX == 1} and \vj{predNotDivX == 0}, which corresponds to
computationally exploit that
$ \Delta_{\mbox{\vj{p}}} $ divides \vj{w}: the \vj{for}-loop body at lines 15--19
becomes accessible, while lines 22--33, with \vj{for}-loops among them,
do not. Lines 15--19 are identical to lines 10--16 of \vj{itF} in
\textbf{Listing}~\ref{reversible core itF} which
we already know to correctly apply \vj{b(x)} and \vj{h(x,y)} in order
to simulate the recursive function we start from.

\medskip
\emph{As a second case},
let $ \mbox{\vj{w\%}}\Delta_{\mbox{\vj{p}}}\ \mbox{\vj{!= 0}} $,
i.e.~$ \Delta_{\mbox{\vj{p}}} $ does not divide the initial value of \vj{x}
which \vj{w} stores in it.
So, $ \mbox{\vj{0}} \not\in I(\mbox{\vj{x}}) $, implying that:
\begin{align}
\label{align:egs non divide}
	\mbox{\vj{e == 0}}
	&&
	\mbox{\vj{g == -}}
	\left\lfloor\frac{\mbox{\vj{w}}}{\Delta_{\mbox{\vj{p}}}}\right\rfloor
	&&
	\mbox{\vj{s == (w\+1)-g-e}}
\end{align}
\noindent
hold at line 9. Lines 11--12 cannot execute, leaving \vj{predDivX} and \vj{predNotDivX} as they are: lines 22--33 become accessible and the \vj{for}-loop at lines 13--19 does not.
\par\noindent
Line 22 increments \vj{w} to balance the information loss that the
rounding of \vj{g} in \eqref{align:egs non divide} introduces;
line 33 recovers the value of \vj{w} when the outer \vj{for}-loop
starts.

\par\noindent
The \vj{if}-selection at lines 25--32 allows to identify when
to apply \vj{b(x)}, which must be followed by the required applications
of \vj{h(x,y)}. We know that $ \mbox{\vj{0}} \not\in I(\mbox{\vj{x}}) $,
so \vj{x == 0} can never hold.
Clearly, \vj{s--} is executed until \vj{x > 0}.
But the \emph{first} time \vj{x > 0} holds true we must compute \vj{b(p(x))},
because the \emph{base} function \vj{b(x)} \emph{must be used the last time} \vj{x} assumes a negative value, \emph{not the first time} it gets positive;
lines 26--30 implement our needs.
Whenever \vj{x > 0} is true, the value of \vj{x} is one step ahead the
one we need: we get one step back with line 26 and,
if it is the first time we step back, i.e.~\vj{z == 0} holds,
then we must execute line 28. If not, i.e.~\vj{z != 0},
we must apply the  \emph{step} function at line 29.
Line 30, restores the right value of \vj{x}.

Finally, we observe that \vj{itG} does not reset \vj{z} to its initial
value \vj{0}. The lines:
\begin{lstlisting}[
language = java,
style = java-style,
caption = Code to set \vj{z} back to \vj{0}.,
label = setting z to 0 again
]
 for (i = 0; i < predNotDivX; i++)  {
   z--;                             }
\end{lstlisting}
which we could place before current line 34, are missing.
The reason not to introduce them is that we want to proceed according
to the initial scheme \eqref{align:compilation scheme}.
We identify a purely reversible part \vj{itGRev}
and a classical part \vj{itGCls} in \vj{itG} which interact,
with no need of \vj{z} anymore. The reason is that
we delegate the control over which between
\vj{b(x)} and \vj{h(x,y)} to apply to the classical part \vj{itGCls}.
This results from framing \vj{itGRev} and \vj{itGCls} as two \textsf{JAVA} threads that synchronously collaborate, under a Producer/Consumer template that \textbf{Section}~\ref{section:Iteration as Producer/Consumer}
introduces.


\section{Reversible vs. Classical as Producer/Consumer}
\label{section:Iteration as Producer/Consumer}
This section comments parts of the \textsf{JAVA} code in \cite{JAVAcode},
that compile and run with
 \href{https://www.oracle.com/java/technologies/javase/jdk14-archive-downloads.html}{\textsf{JAVA SE 14 - ORACLE}}.
The section is meant to be self-contained; an intuitive idea about what a (\textsf{JAVA}) thread should be more than sufficient
to catch the essential points. Endless literature on how using \textsf{JAVA} threads
in details exists and \cite{JAVAcode} is based on one of those possible ``infinite'' instances.

Our goal is to describe how \textsf{JAVA} classes in
\cite{JAVAcode} implement independent threads which, synchronously
collaborating in accordance with a Producer/Consumer template
by means of a couple of communication channels, \emph{fully implement}
both \vj{itGCls} in \textbf{Listing}~\ref{loop on the classical side},
and \vj{itGRev} in \textbf{Listing}~\ref{loop on the reversible side}
in order implement any recursive function traceable to \vj{recG} in \textbf{Listing}~\ref{recG}.
The classes are \vj{ItGCls} (\textbf{Listing~\ref{ItGCls}}),
and \vj{ItGRev} (\textbf{Listing}~\ref{ItGRev}); they define methods
\vj{ItGCls.itGCls} and \vj{ItGRev.itGRev}, respectively.
In particular, \vj{ItGRev.itGRev} fully implements
the purely reversible part of \vj{itG} in \textbf{Listing}~\ref{iteratve recG}
which, we recall, extends the pattern that we find in \textbf{Listing}~\ref{loop on the reversible side}, meant for introductory purposes, and corresponding to
\vj{recF} in \textbf{Listing}~\ref{primitive-recursive-function}, simpler than \vj{recG}.

\begin{figure}
\begin{lstlisting}[
language = java,
style = java-style,
caption = The consumer class \vj{ItGCls}.,
label = ItGCls
]
  public class ItGCls                                  {
    private final Inject inject;
    private final Probe probe;
    private int out = 0;
    private int in = 0;
    public ItGCls(Inject inject, Probe probe, int x) {
      this.inject = inject;
      this.probe = probe;
      this.in = x;                                   }
    public int getOut() { // Let out be available outside ItGCls
      return this.out;  }
    public void itFCls() throws InterruptedException {
      inject.put(in);
      int iterations = probe.get();
      out = B.b(probe.get());
      for (int i = 0; i < iterations; i++) {
        out = H.h(probe.get(), out);     }       }    }
\end{lstlisting}
\end{figure}

As a global assumption, \vj{B} and \vj{H} are two
\textsf{JAVA} classes with (\vj{static}) methods \vj{b(x)} and \vj{h(x,y)}, respectively,
one implementing a \emph{base} function, the other a \emph{step}
function. For example, downloading them from \cite{JAVAcode},
one finds a \vj{B.b(x)} which is the identity, and
a \vj{H.h(x,y)} that returns \vj{x\+y}.

As already outlined, \vj{ItGCls} in \textbf{Listing}~\ref{ItGCls} details out the classical side of \vj{itG}, coherently with \textbf{Listing}~\ref{loop on the classical side}.
Lines 15--17 of \vj{ItGCls} are a classical iteration which,
given a number of \vj{iterations}, and the right sequence of values in the
argument variable \vj{in}, executes the sequence of assignments:
\begin{eqnarray}
\nonumber
&\mbox{\vj{out=B.b(probe.get())}}\\
\label{align:sequence assignments}
&\mbox{\vj{out=H.h(probe.get(),out);}}\\
\nonumber
&\mbox{\vj{out=H.h(probe.get(),out);}}\\
\nonumber
&\mbox{\vj{...}}
\end{eqnarray}
until \vj{out} contains the result, i.e.~what we call \vj{y} in \textbf{Listing}~\ref{loop on the classical side}.
At lines 7--9, \vj{ItGCls} sets the instances \vj{inject} and \vj{probe}
of the two communication channels \vj{Probe} and \vj{Inject}.
Lines 15, 17 obtain the required argument values of the
\emph{base} and of the \emph{step} functions by calling \vj{probe.get()}.
Also line 14 calls \vj{probe.get()}. It is to obtain the number of \vj{iterations}. Instead, line 13 sends the initial value of \vj{in}
to the reversible side, i.e.~(an instance of) the producer \vj{ItGRev};
\vj{ItGRev} needs such a value
to start producing the sequence of arguments values that \vj{ItGCls}
requires to execute the sequence~\eqref{align:sequence assignments}.

\begin{figure}
\begin{lstlisting}[
language = java,
style = java-style,
caption = The (reversible) producer \vj{ItGRev}.,
label = ItGRev
]
 public class ItGRev                                      {
   private final Inject inject;
   private final Probe probe;
   ItGRev(Inject inject, Probe probe) {
     this.inject = inject;
     this.probe = probe;            }
   public void itGRev() throws InterruptedException      {
     int s = 0, e = 0, g = 0, w = 0, x = 0;
     int predDivX = 0, predNotDivX = 1;
     x = inject.swapIn(x); // read x from itGCls
     w = w + x;
     for (int i = 0; i <= w; i++)  {
       if      (x >  0) { g++; }
       else if (x == 0) { e++; }
       else             { s++; }
       x = Pred.pred(x);           }
     for (int i = 0; i < e; i++)               {
       predDivX = predDivX + predNotDivX;
       predNotDivX = predDivX - predNotDivX;   }
     for (int j = 0; j < predDivX; j++)      {
       probe.put(g); // send g to itGCls for iterations
       for (int i = 0; i <= w; i++)         {
         x = Pred.predInv(x);
         if      (x >  0) { g--;
                            // send x to itGCls for h(x,y)
                            probe.put(x); }
         else if (x == 0) { e--;
                            // send x to itGCls for b(x)
                            probe.put(x); }
         else             { s--;          }}}
     for (int j = 0; j < predNotDivX; j++)         {
       probe.put(g); // send g to itGCls for iterations
       w++;
       for (int i = 0; i <= w; i++)              {
         x = Pred.predInv(x);
         if      (x >  0) { g--;
                            x = Pred.pred(x);
                            // send x to itGCls for b(x) or h(x,y)
                            probe.put(x);
                            x = Pred.predInv(x);}
         else if (x == 0) { e--;                }
         else             { s--;                }}
       w--;                                       }
     w = w - x;
     x = inject.swapOut(x); // restore initial x  }}
\end{lstlisting}
\end{figure}

As told, \textbf{Listing}~\ref{ItGRev} is a possible implementation of \textbf{Listing}~\ref{iteratve recG}
up to the parts that make \textbf{Listing}~\ref{iteratve recG} irreversible.
The table:
\begin{center}
\begin{tabular}{c|c}
 \mbox{\vj{itG}}     & \mbox{\vj{ItGRev}} \\\hline\hline
 4--8 \quad   & 12--16      \\\hline
10--12\quad   & 17--19      \\\hline
14--19\quad   & 20--30      \\\hline
21--33\quad   & 31--43
\end{tabular}
\end{center}
list code lines in \vj{itG}
of \textbf{Listing}~\ref{iteratve recG} that correspond in \vj{ItGRev} of
\textbf{Listing}~\ref{ItGRev}.

We now trace the behavior of \vj{ItGRev} in analogy to the one of \vj{itG}.
Let $ \mbox{\vj{w\%}}\Delta_{\mbox{\vj{p}}}\mbox{\vj{== 0}} $, so lines 21--30 are accessible.
Line 21 sets the value of \vj{iterations} in
the consumer \vj{ItGCls}, so it can move to its line 15
waiting for the argument value of \vj{B.b(x)}.
Lines 26, and 29 call \vj{probe.put()}; in both cases
the call sets the value that the consumer
\vj{ItGCls} waits in order to feed \vj{H.h(x,out)} or \vj{B.b(x)}.

Let, instead, $ \mbox{\vj{w\%}}\Delta_{\mbox{\vj{p}}}\mbox{\vj{!= 0}} $,
so lines 32--43 are accessible.
Line 32 sets the value of \vj{iterations} in
the consumer \vj{ItGCls} exactly like line 21.
Lines 36--42 simplify lines 25--32 of \vj{itG} in
\textbf{Listing}~\ref{iteratve recG}, for \emph{they do not need}
neither an \vj{if}-selection nor \vj{z} to identify which between
\vj{B.b(x)} and \vj{H.h(x,out)} to use, operation delegated to \vj{ItGCls}
which receives \vj{x} from \vj{probe.put(x)} at line 39.

Finally, the producer \vj{ItGRev} is triggered to start at line 10:
\vj{inject.swapIn()} swaps the content of \vj{x}, ancilla local to
\vj{ItGRev}, with the input value, (possibly) available in \vj{Inject.inject},
shared between \vj{ItFRev} and \vj{ItGCls}.
Line 45 restores the initial value of \vj{x} by swapping it with the one
it gets at line 10.

\begin{figure}
\begin{lstlisting}[
language = java,
style = java-style,
caption = Channel \vj{Probe}.,
label = Probe
]
  public class Probe                                             {
  	private int x = 0;
  	private boolean xAvailable = false;
  	public synchronized int get()
  	         throws InterruptedException                {
  		while (!xAvailable) { wait(); } // producer has not produced
  		int out = x;
  		xAvailable = !xAvailable;
  		notify();
  		return out;                                             }
  	public synchronized void put(int in)
  	         throws InterruptedException                   {
  		while (xAvailable) { wait(); } // consumer has not consumed
  		x = in;
  		xAvailable = !xAvailable;
  		notify();                                                 }}
\end{lstlisting}
\end{figure}

Instances of classes \vj{Probe} and \vj{Inject} model channels
for synchronous dialogues between (instances of) \vj{ItGCls} and \vj{ItGRev}.
\textsf{JAVA} `\vj{synchronized}' directive let the bodies
of every method in \vj{Probe} and \vj{Inject} be critical regions,
each one accessed by a single thread at a time, one running \vj{ItGCls}, the other \vj{ItGRev}.
Let us comment on method \vj{get} in \vj{Probe}, all others, those ones of
\vj{Inject} included, behaving analogously
\footnote{We address to the official documentation on \textsf{JAVA} classes \vj{Runnable} and \vj{Threads} on possible implementations of critical regions.}.
Let \vj{probe.get()} be a call in \vj{ItGCls} of an instance \vj{probe} of \vj{Probe}.
The consumer enters \vj{probe.get()} and \vj{waits()} until
\vj{xAvailable} is set to \vj{true}, meaning that the producer \vj{ItGRev}
made a value available in \vj{probe.x} by means of a call
to \vj{probe.put(x)}. As soon as \vj{ItGRev} negates \vj{xAvailable},
setting it to \vj{true}, and calls \vj{notify()} at line 16 in
\vj{probe.put(x)}, \vj{ItGCls} leaves line 6 of \vj{probe.get()} in order to:
(i) set \vj{out} with the value in \vj{x};
(ii) negate \vj{xAvailable} again;
(iii) notify that \vj{ItGRev} can produce a new value for \vj{probe.x};
(iv) return the value of \vj{probe.out}.
Method \vj{prob.put(x)} sets a symmetric behavior on \vj{ItGRev}.

\begin{figure}
\begin{lstlisting}[
language = java,
style = java-style,
caption = Channel \vj{Inject}.,
label = Inject
]
  public class Inject                         {
    private int xInitial = 0;
    private boolean notSet = true;
    public synchronized int get()
            throws InterruptedException      {
      while (notSet) { wait(); }
      int out = xInitial;
      notify();
      return out;                            }
    public synchronized void put(int in)
             throws InterruptedException     {
      while (!notSet) { wait(); }
      xInitial = in;
      notSet = !notSet;
      notify();                              }
    public synchronized int swapIn(int in)
            throws InterruptedException      {
      while (notSet) { wait(); }
      int out = xInitial;
      xInitial = in;
      notify();
      return out;                            }
    public synchronized int swapOut(int in)
             throws InterruptedException     {
      int out = xInitial;
      xInitial = in;
      notSet = !notSet;
      notify();
      return out;                            }}
\end{lstlisting}
\end{figure}


\section{Conclusions, related and future work}
\label{section:Conclusions}
In this work we show that given a (unary) \emph{base} function \vj{b(x)},
a (binary) \emph{step} function \vj{h(x,y)}, a
(unary)  \emph{predecessor} \vj{p(x)}
that decreases every of its input \vj{x} by a constant value
$ \Delta_{\mbox{\vj{p}}} $, so the inverse \vj{pInv(x)} forcefully exists,
we can decompose every recursive (classical) function
\vj{recG[p,b,h]}, built on \vj{p(x)}, \vj{b(x)}, and
\vj{h(x,y)} as in \textbf{Listing}~\ref{recG} into two iterative sides:
the reversible \vj{itGRev[p,pInv]}, and the classical \vj{itGCls[b,h]},
following the notation in \eqref{align:compilation scheme}.
Both \vj{itGRev[p,pInv]} and \vj{itGCls[b,h]} synchronously cooperate
to implement \vj{recG[p,b,h]} as Producer/Consumer.
Despite the simplicity of \vj{recG[p,b,h]},
its decomposition requires some work in order to soundly model the recursive
unfolding process by means of finite iterations available in intrinsically reversible programming formalism like \textsf{SRL}, and \textsf{RPP}
which \vj{itGRev[p,pInv]} can be traced to.
The decomposition into two parts finds implementation as \textsf{JAVA}
threads which fit both Perumalla's ``Source-to-Source Translator'' and ``Library-based'' approaches to reversibility.

The transformation we propose shows the role that reversibility
can have in everyday programming.
Let \vj{recG} be matching \textbf{Listing}~\ref{recG}
for a possible example scenario which follows.
Being \vj{recG} recursive, typically it is simpler to prove correct, as compared to an equivalent iterative one.
A compiler should generate \vj{ItGCls} and \vj{ItGRev} equivalent to
\vj{recG}, such that (istances of) \vj{ItGRev} could run on a true reversible
hardware, possibly contributing to leverage the promised greener foot-print
of reversible computing, as compared to the classical one.

A first natural step beyond this work is to identify a class \textsf{R}
of recursive schemes more general than the one in \textbf{Listing}~\ref{recG}.
Of course, \textsf{R} must contain recursive functions with arbitrary arity.
Second, the recursive functions of \textsf{R} should be able to be based on  \emph{predecessors} \vj{p} such that, at least:
\begin{enumerate}
	\item
	$ \Delta_{\mbox{\vj{p}}} $ is not necessarily a constant, as
	in \textbf{Section}~\ref{section:From recursion to iteration}.
    For example, $ \Delta_{\mbox{\vj{p}}}\ \mbox{\vj{== -3}}$ on even arguments, and \vj{-2} on odd ones can be useful;
	\item
	\vj{p(x)} is an integer division \vj{x/k}, for some given \vj{k > 0},
	like in a dichotomic search, that has \vj{k == 2}.
\end{enumerate}
More generally, we aim at an \textsf{R} with recursive functions only,
of which every \vj{recG} is defined
on at least one \emph{predecessors} \vj{p} such that, for every argument \vj{x},
a finite iterated application \vj{p(..p(x)..))} exists which let the \emph{condition} \vj{c(p(..p(x)..)} of \vj{recG} be true.
So, for every \vj{recG[p,b,h]}, \vj{p}, \vj{b} and \vj{h} in \textsf{R}
the introductory scheme \eqref{align:compilation scheme}
would become a compilation scheme $ \llbracket \cdot \rrbracket $
from \textsf{R} to, just as an example, \textsf{JAVA} threads
``defined'' as:
\begin{align*}
	\llbracket \mbox{\vj{p}} \rrbracket & = \mbox{\vj{required (reversible) code}}
	\\
	\llbracket \mbox{\vj{pInv}} \rrbracket & =
   \mbox{\vj{!}}\llbracket \mbox{\vj{p}} \rrbracket
    \\
	\llbracket \mbox{\vj{recG[p,b,h]}} \rrbracket & =
	\mbox{\vj{ItGCls[$\llbracket$b$\rrbracket$,$\llbracket$h$\rrbracket$]}}
	\parallel \mbox{\vj{ItGRev[$\llbracket$p$\rrbracket$,$\llbracket$pInv$\rrbracket$]}}
	\enspace,
\end{align*}
where
$ \mbox{\vj{!}}\llbracket . \rrbracket $ inverts the code that its
argument $ \llbracket . \rrbracket $ produces, and
$ \parallel $ models some kind communication.

Also, we see the above compilation scheme
a good starting point to abstract away from the concreteness oriented
perspective taken so far in this work. The new goal would be
to investigate if the decomposition
we have seen has some analogies with Girard's decomposition
$ A \rightarrow B \simeq\ !A \multimap\ B $.
To us, decomposing \vj{recG[p,b,h]} in terms of
\vj{itGCls[b,h]} and \vj{itGRev[p,pInv]} suggests that the relation between
reversible and classical computations can be formalized by
a linear isomorphism $ A^n \multimapboth B^n$ between tensor products
$ A^n$, and $B^n $ of $ A$, and $B $, in analogy to  \cite{DBLP:conf/popl/JamesS12}; then we can get classical computations by
applying a functor, say $ \gamma $, whose purpose is, at least, to forget, or to inject replicas of, parts
of $ A^n$, and $B^n $, in a way that
$ (\gamma A^n \rightarrow \gamma A^n) \uplus
  (\gamma A^n \leftarrow  \gamma A^n)  $
can be a proposal for their type;
the type says that we pass from a reversible computation to a classical one
by choosing which is input and which is output, and by introducing freedom in the use of the computational resources.

In the same foundational vein, we conclude by observing
that the decomposition of a given \vj{revG} that this work introduces, can
possibly lead to identify a hierarchy inside the class \textsf{R}, once
\textsf{R} is precisely identified.
The hierarchy would characterize a function \vj{revG} depending on the
computational space that the reversible component \vj{itGRev} we would
extract from \vj{revG} requires to work; intuitively, the space would
depend on how complex the inverse \vj{pInv} of the \emph{predecessor}
\vj{p} is, and that \vj{revG} is defined on.
We see this as related to
\cite{DBLP:journals/corr/abs-cs-0504088-Vitanyi,DBLP:journals/mvl/YokoyamaAG12}.


\bibliographystyle{plain}
\bibliography{bib-minimal}

\begin{thebibliography}{10}

\bibitem{JAVAcode}
Eclipse java project rev2iterrev.
\newblock \url{https://github.com/LucaRoversi/Rec2IterRev}.

\bibitem{BOITEN1992139}
Eerke~A. Boiten.
\newblock Improving recursive functions by inverting the order of evaluation.
\newblock {\em Science of Computer Programming}, 18(2):139 -- 179, 1992.

\bibitem{GIRARD19871}
Jean-Yves Girard.
\newblock Linear logic.
\newblock {\em Theoretical Computer Science}, 50(1):1 -- 101, 1987.

\bibitem{Gries1981}
David Gries.
\newblock {\em The Science of Programming}.
\newblock Texts and Monographs in Computer Science. Springer, New York, NY.

\bibitem{DBLP:conf/popl/JamesS12}
Roshan~P. James and Amr Sabry.
\newblock Information effects.
\newblock In John Field and Michael Hicks, editors, {\em Proceedings of the
  39th {ACM} {SIGPLAN-SIGACT} Symposium on Principles of Programming Languages,
  {POPL} 2012, Philadelphia, Pennsylvania, USA, January 22-28, 2012}, pages
  73--84. {ACM}, 2012.

\bibitem{DBLP:journals/tcs/Matos03}
Armando~B. Matos.
\newblock Linear programs in a simple reversible language.
\newblock {\em Theor. Comput. Sci.}, 290(3):2063--2074, 2003.

\bibitem{MatosRC2020}
Armando~B. Matos, Luca Paolini, and Luca Roversi.
\newblock On the expressivity of total reversible programming languages.
\newblock In Ivan Lanese and Mariusz Rawski, editors, {\em Reversible
  Computation}, pages 128--143, Cham, 2020. Springer International Publishing.

\bibitem{paolini2017ngc}
Luca Paolini, Mauro Piccolo, and Luca Roversi.
\newblock On a class of reversible primitive recursive functions and its
  turing-complete extensions.
\newblock {\em New Generation Computing}, 36(3):233--256, Jul 2018.

\bibitem{DBLP:journals/tcs/PaoliniPR20}
Luca Paolini, Mauro Piccolo, and Luca Roversi.
\newblock A class of recursive permutations which is primitive recursive
  complete.
\newblock {\em Theor. Comput. Sci.}, 813:218--233, 2020.

\bibitem{perumalla2013chc}
Kalyan~S. Perumalla.
\newblock {\em Introduction to Reversible Computing}.
\newblock Chapman \& Hall/CRC Computational Science. Taylor \& Francis, 2013.

\bibitem{10.1093/comjnl/bxm116}
J.~E. Rice.
\newblock {An Introduction to Reversible Latches}.
\newblock {\em The Computer Journal}, 51(6):700--709, 01 2008.

\bibitem{DBLP:journals/corr/abs-cs-0504088-Vitanyi}
Paul M.~B. Vit{\'{a}}nyi.
\newblock Time, space, and energy in reversible computing.
\newblock {\em CoRR}, abs/cs/0504088, 2005.

\bibitem{DBLP:journals/mvl/YokoyamaAG12}
Tetsuo Yokoyama, Holger~Bock Axelsen, and Robert Gl{\"{u}}ck.
\newblock Optimizing reversible simulation of injective functions.
\newblock {\em J. Multiple Valued Log. Soft Comput.}, 18(1):5--24, 2012.

\end{thebibliography}


\nocite{10.1093/comjnl/bxm116}
\nocite{Gries1981}

\end{document}